\documentclass[a4paper]{jpconf}
\usepackage{graphicx}
\usepackage{aligned-overset}
\usepackage{amsfonts}

\begin{document}
\title{Unconventional singlet-triplet superconductivity}

\author{David Möckli}

\address{Instituto de F\'{i}sica, Universidade Federal do Rio Grande do Sul, 91501-970 Porto Alegre, Brazil}

\ead{mockli@ufrgs.br}

\begin{abstract}
Have you been lying awake wondering what symmetries determine whether a superconductor is spin singlet, triplet, or both? 
We show that if BCS theory is supplied with additional degrees of freedom, spin singlet can coexist with spin triplet superconductivity. 
In the first part, we didactically guide the reader to the most general superconducting state using symmetry arguments. 
If both singlet and triplet pairing channels are present, a magnetic field can convert between spin singlet and triplet states. 
In the second part,
we discuss two possible singlet-triplet superconductors: CeRh$_2$As$_2$ and bilayer-NbSe$_2$.
\end{abstract}

\section{Introduction}

The two key symmetries in conventional Bardeen-Cooper-Schrieffer (BCS) theory are time-reversal and inversion \cite{sigrist_unc,sigrist_tutorial}. The BCS superconducting ground state $|\psi_\mathsf{BCS}\rangle$ has a rigid phase, which occurs through the $U(1)$ phase symmetry breaking of the normal state. 
Unconventional superconductors break additional symmetries such as rotation, or the key symmetries (inversion and/or time-reversal) of BCS theory. 
Another venue to unconventional superconducting states is examining additional degrees of freedom (DOFs), such as orbital, valley, sublattice, etc., which introduce non-trivial symmetries. 
Here, we introduce the superconducting state-vector of a superconductor with additional DOFs. 
The additional DOF naturally leads to the possibility of having a superposition of singlet and triplet components in the state-vector, in contrast to non-centrosymmetric superconductors where the explicit removal of inversion symmetry is required. 

It might be hard to measure the triplet component in singlet-triplet superconductors, since a singlet $s$-wave state gaps out the Fermi surface \cite{menashe2020}.  Yet, if triplet components in singlet-triplet mixed superconductors turn out to exist, this might open an alternative option to access the properties of triplet superconductivity. Up to now, triplet superconductivity is limited only to a few pure triplet superconductors in Uranium based compounds. We introduce the conditions for singlet-triplet mixing and discuss possible applications to CeRh$_2$As$_2$ and bilayer NbSe$_2$.

\section{The superconducting state-vector}

Consider a superconducting order parameter $\Delta_{s_1s_2}^{ab}(\mathbf{k})\propto \langle c_{-\mathbf{k}s_2b}c_{\mathbf{k}s_1a}\rangle $, where $\mathbf{k}$ is momentum, $s_1,s_2=\{\uparrow,\downarrow\}$ is the $z$-spin projection, and $a,b=\{1,2\}$ is an additional internal DOF that could refer to orbitals, sublattice or layers.
We assume that corresponding superconducting state-vectors to the order parameter can be written as (omitting an overall phase and normalization factor)
\begin{equation}
    \left|\psi_{sm}^{ab}(\mathbf{k})\right\rangle =\phi(\mathbf{k})|\chi_{sm}\rangle\otimes |\Upsilon_{ab}\rangle,
\end{equation}
where $s$ is the total spin and $m=s_1+s_2$, which is $s=m=0$ for a singlet state and $s=1$ for the triplet states $m=\{1,0,-1\}$ \cite{sakurai_napolitano_2017}. Let us write the singlet state as $|\psi_{S}^{ab}(\mathbf{k})\rangle =\phi(\mathbf{k})|\chi_{00}\rangle\otimes |\Upsilon_{ab}\rangle = \phi(\mathbf{k})(|\uparrow\downarrow\rangle-|\downarrow\uparrow\rangle)|\otimes |\Upsilon_{ab}\rangle$ and the triplet states $|\psi_{Tm}^{ab}(\mathbf{k})\rangle =\phi(\mathbf{k})|\chi_{1,m}\rangle\otimes |\Upsilon_{ab}\rangle $, where $|\chi_{1,0}\rangle = |\uparrow\downarrow\rangle+|\downarrow\uparrow\rangle$, $|\chi_{1,1}\rangle = |\uparrow\uparrow\rangle$ and $|\chi_{1,-1}\rangle = |\downarrow\downarrow\rangle$. 
For the three triplets $\phi(\mathbf{k})|\chi_{1m}\rangle$, we also could have used the three component $\mathbf{d}(\mathbf{k})$-vector parametrization \cite{sigrist_unc}, but we stick with $\phi(\mathbf{k})|\chi_{1m}\rangle$ to maintain the discussion accessible to anyone familiar with composite two spin states.
For the four possibilities of $|\Upsilon_{ab}\rangle$ we choose $|\Upsilon_{11}\rangle=|1,1\rangle+|2,2\rangle$, $|\Upsilon_{22}\rangle=|1,1\rangle-|2,2\rangle$, $|\Upsilon_{12}\rangle=|1,2\rangle+|2,1\rangle$ and $|\Upsilon_{21}\rangle=|1,2\rangle-|2,1\rangle$. 
The four possible spin states combined with the four possible internal DOF states gives a total of 16 possible configurations, as opposed to only 4 possibilities without the additional DOF. 

\subsection{Permutation and inversion symmetry}

The Pauli principle enforces that under the action of the permutation operator $\mathsf{P}|\psi_{sm}^{ab}(\mathbf{k})\rangle=-|\psi_{sm}^{ab}(\mathbf{k})\rangle$. 
The permutation operator $\mathsf{P}$ takes $\mathbf{k}\rightarrow -\mathbf{k}$, $s_1\leftrightarrow s_2$, and exchanges the particle subspaces $a\leftrightarrow b$. Then we have for singlets and triplets, respectively
\begin{equation}
\mathsf{P}\left|\psi_S^{ab}(\mathbf{k})\right\rangle =\mathsf{P}\left[ \phi(\mathbf{k})|\chi_{00}\rangle\otimes|\Upsilon_{ab}\rangle\right ]=-\phi(-\mathbf{k})|\chi_{00}\rangle\otimes|\Upsilon_{ba}\rangle\overset{\mbox{\tiny Pauli}}{=} 
-\phi(\mathbf{k})|\chi_{00}\rangle\otimes|\Upsilon_{ab}\rangle;
\label{eq:pauliS} 
\end{equation}
\begin{equation}
\mathsf{P}\left|\psi_{Tm}^{ab}(\mathbf{k})\right\rangle =\mathsf{P}\left[ \phi(\mathbf{k})|\chi_{1,m}\rangle\otimes|\Upsilon_{ab}\rangle\right ]=\phi(-\mathbf{k})|\chi_{1,m}\rangle\otimes|\Upsilon_{ba}\rangle\overset{\mbox{\tiny Pauli}}{=} 
-\phi(\mathbf{k})|\chi_{1,m}\rangle\otimes|\Upsilon_{ab}\rangle .
\label{eq:pauliT}
\end{equation}
If we ignore the internal DOF $|\Upsilon_{ab}\rangle$, the last equality in Eqs. \eqref{eq:pauliS} and \eqref{eq:pauliT} tell us that for a singlet (triplet) state the momentum structure $\phi(\mathbf{k})$ must be even (odd) \cite{sigrist_unc,sigrist_tutorial}. However, in the presence of $|\Upsilon_{ab}\rangle$, even-momentum-triplets and odd-momentum-singlets can now exist, because the Fermionic antisymmetry can now be carried by an additional DOF $|\Upsilon_{ab}\rangle$. 

It is instructive to determine the inversion (parity) sectors of the superconducting state-vectors. The action of the inversion operator $\mathsf{I}\left|\psi_{sm}^{ab}(\mathbf{k})\right\rangle$ takes $\mathbf{k}\rightarrow -\mathbf{k}$, leaves the spins $\{s_1,s_2\}$ invariant, and we assume that $\mathsf{I}|\Upsilon_{ab}\rangle$ exchanges not the particle subspaces, but the DOFs $1\leftrightarrow 2$,
as would be the case when different sublattices are related by inversion. 
Then we have
\begin{equation}
\mathsf{I}\left|\psi_{sm}^{ab}(\mathbf{k})\right\rangle =\mathsf{I}\left[ \phi(\mathbf{k})|\chi_{sm}\rangle\otimes|\Upsilon_{ab}\rangle\right ]=\phi(-\mathbf{k})|\chi_{sm}\rangle\otimes\mathsf{I}|\Upsilon_{ab}\rangle =
\phi(-\mathbf{k})|\chi_{sm}\rangle\otimes
\begin{cases}
|\Upsilon_{ab}\rangle,\, (a=1) \\
-|\Upsilon_{ab}\rangle,\, (a=2).
\end{cases}
\label{eq:inversion}
\end{equation}
If we ignore $|\Upsilon_{ab}\rangle$, Eq. \eqref{eq:inversion} tells us that singlets ($\phi(\mathbf{k})=\phi(-\mathbf{k})$) belong to the $+1$ inversion sector, whereas triplets ($\phi(\mathbf{k})=-\phi(-\mathbf{k})$) belong to the $-1$ inversion sector. 
Therefore, if the system is inversion symmetric, the superconducting state-vector must have a definite parity, that is, it either belongs to the parity eigenvalue sector $+1$ or $-1$. 
In this case, inversion symmetry prohibits a superposition of a singlet with a triplet state.
When an inversion symmetric superconducting material lacks additional DOFs, one then knows that its superconducting state is either singlet or triplet. 
What if inversion symmetry lacks? Then there is no definite parity (and no selection rule) which allows a singlet-triplet superposition in the state-vector. 
The interesting question is then, perhaps: what is the singlet-triplet volume ratio? We return to this later.
Yet, the main message of Eq. \eqref{eq:inversion} is that in the presence of additional DOFs $|\Upsilon_{ab}\rangle$, a singlet-triplet superposition is symmetry allowed even in inversion symmetric systems. 

\subsection{Singlet-triplet states}

\begin{figure}
\includegraphics[width=0.98\textwidth]{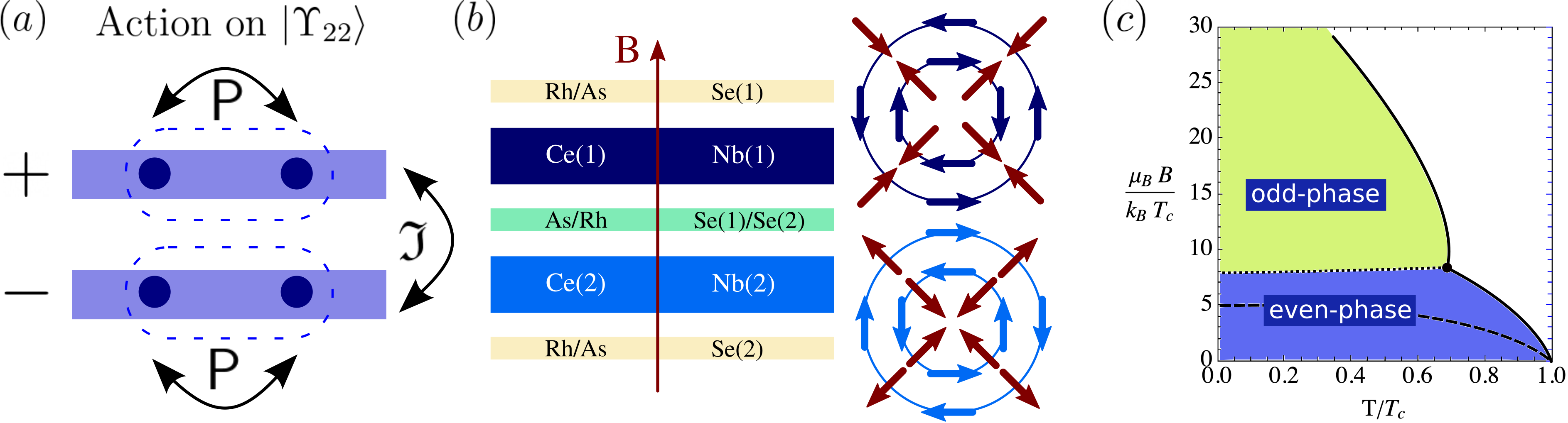}
\caption{\label{fig:1} (a) Action of the permutation $\mathsf{P}$ and inversion $\mathsf{I}$ operators on the sublattice subspace $|\Upsilon_{22}\rangle=|1,1\rangle-|2,2\rangle$. 
$\mathsf{P}|\Upsilon_{22}\rangle=|\Upsilon_{22}\rangle$, but $\mathsf{I}|\Upsilon_{22}\rangle=-|\Upsilon_{22}\rangle$.
(b) Schematic of CeRh$_2$As$_2$ (left) and bilayer NbSe$_2$ (right). The blue arrows show the in-plane Rashba SOC texture of the inversion related Ce(1) and Ce(2) sublattices. The red arrows show the texture of the induced triplet $d$-vector by a perpendicular magnetic field $\mathbf{B}$. We do not show the textures for NbSe$_2$, which would have an Ising SOC in the $z$ direction, and an induced $d$-vector texture for an in-plane magnetic field.
The Nb$(n)$ layers have a local mirror symmetry, which is why the SOC is of Ising type.
(c) Phase diagram for CeRh$_2$As$_2$ under a perpendicular magnetic field. The solid line indicates a second-order phase transition. The dotted-line is the first order transition between the even and odd phases. The dashed-line is the upper critical field for an in-plane field.
The same phase diagram can be reinterpreted for bilayer NbSe$_2$ under an in-plane field.}
\end{figure}

As a concrete example, consider the superposition $|\psi(\mathbf{k})\rangle=c_1|\psi_S^{11}(\mathbf{k})\rangle+c_2|\psi_{T0}^{22}(\mathbf{k})\rangle $, where $c_1$ and $c_2$ are complex coefficients. By applying the permutation operator, we obtain
\begin{equation}
\mathsf{P}|\psi(\mathbf{k})\rangle=c_1\mathsf{P}|\psi_S^{11}(\mathbf{k})\rangle+c_2\mathsf{P}|\psi_{T0}^{22}(\mathbf{k})\rangle 
=c_1\phi_1(-\mathbf{k})\left[-|\chi_{00}\rangle \right ]\otimes |\Upsilon_{11}\rangle+
c_2\phi_2(-\mathbf{k})|\chi_{T,0}\rangle\otimes |\Upsilon_{22}\rangle. 
\end{equation}
The Pauli principle then imposes that $\phi_1(\mathbf{k})$ is an even function and $\phi_2(\mathbf{k})$ must be odd. With this knowledge, we now apply the inversion operator to check the parity sectors. 
We obtain $\mathsf{P}|\psi_S^{11}(\mathbf{k})\rangle=(+1)|\psi_S^{11}(\mathbf{k})\rangle$ and $\mathsf{P}|\psi_{T0}^{22}(\mathbf{k})\rangle=(+1)|\psi_{T0}^{22}(\mathbf{k})\rangle$, where we used $\mathsf{I}|\Upsilon_{22}\rangle=-|\Upsilon_{22}\rangle$. 
Therefore, because of the additional DOF, $|\psi_S^{11}(\mathbf{k})\rangle$ singlets and $|\psi_{T0}^{22}(\mathbf{k})\rangle$ triplets belong to the same parity, which allows them to coexist in a superposition even in an inversion symmetric material. 
The mathematical subtlety relies on the fact that $\mathsf{P}|\Upsilon_{22}\rangle=|\Upsilon_{22}\rangle$, but $\mathsf{I}|\Upsilon_{22}\rangle=-|\Upsilon_{22}\rangle$; see Fig. \ref{fig:1}(a). 

The most general even and odd superpositions are
\begin{equation}
|\psi_\mathsf{even}\rangle  =  c_1|\psi_S^{11}\rangle+c_2|\psi_S^{12}\rangle+c_3|\psi_S^{21}\rangle
+c_4|\psi_{T0}^{22}\rangle+c_5|\psi_{T,1}^{22}\rangle+c_6|\psi_{T,-1}^{22}\rangle; \label{eq:even}
\end{equation}
\begin{equation}
\begin{split}
 |\psi_\mathsf{odd}\rangle  &  =  c_7|\psi_S^{22}\rangle +c_8|\psi_{T0}^{11}\rangle +c_9|\psi_{T0}^{12}\rangle+c_{10}|\psi_{T0}^{21}\rangle
+c_{11}|\psi_{T1}^{11}\rangle+c_{12}|\psi_{T,-1}^{11}\rangle
+c_{13}|\psi_{T1}^{12}\rangle+c_{14}|\psi_{T,-1}^{12}\rangle \\
& +c_{15}|\psi_{T1}^{21}\rangle+c_{16}|\psi_{T,-1}^{21}\rangle. 
\end{split} \label{eq:odd}
\end{equation}
Here, all the 16 possible states are assigned to an even or odd sector. If the system lacks inversion, then $|\Psi\rangle=|\psi_\mathsf{even}\rangle+|\psi_\mathsf{odd}\rangle$. 
The singlet state $c_7|\psi_S^{22}\rangle$ deserves special attention, since it is the only parity-odd-singlet. Suppose that the pairing mechanism of an inversion symmetric material only admits singlet instabilities. This means that because of the additional DOF, there are always mutually excluding even-singlet and odd-singlet instabilities that might \textit{switch} depending on external parameters.
This is likely to be the case in CeRh$_2$As$_2$ \cite{khim2021fieldinduced}, and possibly also in few-layer NbSe$_2$ \cite{kuzmanovic}. 

\section{Effect of a Zeeman field}

What are the interesting properties of the perhaps so far speculated singlet-tripled mixed superconductors? 
One interesting aspect is how the multicomponent superconducting state responds to a Zeeman magnetic field. It is intuitive to understand why. Heuristically, a Zeeman field applied to a singlet state $|\chi_{00}\rangle=|\uparrow\downarrow\rangle-|\downarrow\uparrow\rangle$ will try to polarize the spins along the magnetic field and thus breaking up the singlet configuration. 
In contrast, a Zeeman field applied for instance to the triplet state $|\chi_{1,\pm 1}\rangle$ only favors it. 
Even more, if there is an attractive pairing in both singlet and triplet channels, instead of breaking up the singlets, the Zeeman field can convert the singlet configuration to a triplet. Then, the field can control the singlet-triplet volume ratio. 

Let us give a simple geometric description of the singlet-triplet conversion process. Consider a 2D superconductor with the quantization axis set perpendicularly to the 2D plane. 
The superconducting state at zero field is taken to be the pure singlet \begin{equation}
 |\Psi_0(\mathbf{k})\rangle =|\psi_S^{11}(\mathbf{k})\rangle =\phi(\mathbf{k})|\chi_{00}\rangle\otimes |\Upsilon_{11}\rangle= \left[|\mathbf{k}\uparrow\rangle|-\mathbf{k}\downarrow\rangle -|\mathbf{k\downarrow\rangle|-\mathbf{k}\uparrow}\rangle\right ]\otimes |\Upsilon_{11}\rangle.
\end{equation}
We now apply a magnetic field along the $y$-axis. In the very large magnetic field limit, according to the heuristic conversion argument, we expect all singlets to convert to triplets. The action of the $y$-directed magnetic field on a $|\mathbf{k}\uparrow\rangle$ state can be implemented as a spin rotation around the $x$-axis $\mathsf{R}_x(\theta)=\exp(-i\sigma_x\theta/2)$. 
In the large magnetic field limit, and taking into account that Cooper pairing occurs at opposite momenta, the action of the magnetic field $\mathsf{B}_y$ on an arbitrary state is $\mathsf{B}_y=\mathsf{R}_x\left[\mathrm{sgn}(\pm\mathbf{k})\pi/2 \right ]$, where the sign function determines whether the state is located at positive or negative momenta \cite{PhysRevB.99.180505}. Since $\mathsf{B}_y$ does not act on $|\Upsilon_{11}\rangle$, we omit it in the following. We have
\small
\begin{equation}
\mathsf{B}_y |\Psi_0(\mathbf{k})\rangle =R_x\left(\frac{\pi}{2} \right )|\mathbf{k}\uparrow\rangle 
R_x\left(-\frac{\pi}{2} \right )|-\mathbf{k}\downarrow\rangle -R_x\left(\frac{\pi}{2} \right )|\mathbf{k}\downarrow\rangle 
R_x\left(-\frac{\pi}{2} \right )|-\mathbf{k}\uparrow\rangle =i\left(|\psi_{T1}(\mathbf{k})\rangle-|\psi_{T,-1}(\mathbf{k})\rangle \right ).
\label{eq:conversion}
\end{equation}
\normalsize
The action of $\mathsf{B}_y$ converted $|\psi_S(\mathbf{k})\rangle$ to $\pm i|\psi_{T,\pm 1}(\mathbf{k})\rangle$. 
Such a Zeeman induced singlet to triplet conversion can occur in inversion symmetric superconductors with additional DOFs \cite{mockli2021}  or in non-centrosymmetric materials where no additional DOF is necessary \cite{PhysRevB.99.180505,ising_eilenberger_2020,mockli2021}.
An interesting theoretical problem from magnetic field-induced singlet to triplet conversion that could develop from this is considering a singlet-triplet mixed Abrikosov vortex lattice and studying how the conversion mechanism affects it. 

\section{The pairing interaction}

Up to now, we presented step by step calculations. From now on, the mathematics will be pictorial for discussion purposes. Omitting the additional DOFs for simplicity, a superconducting pairing interaction has the form
\begin{equation}
\mathsf{H}_\mathsf{int}  = \frac{1}{2}\sum_{\mathbf{k},\mathbf{k}'}\sum_{\{s_i\}}V_{s_1s_2,s_1's_2'}(\mathbf{k},\mathbf{k}')c_{\mathbf{k}s_1}^\dag c_{-\mathbf{k}s_2}^\dag c_{-\mathbf{k}'s_2'}c_{\mathbf{k}'s_1}.
\end{equation}
The present interaction neglects the possibility of finite momentum pairing (which is an interesting line of investigation). 
The pairing interaction can be decomposed into singlet and triplet channels as \cite{Frigeri2006}
\scriptsize
\begin{equation}
V_{s_1s_2,s_1's_2'}(\mathbf{k},\mathbf{k}')=\sum_{\Gamma,j}\left(v_{s,\Gamma}\left[\hat{\psi}_{\Gamma_j}(\mathbf{k})i\sigma_2 \right ]_{s_1s_2}\left[\hat{\psi}_{\Gamma_j}(\mathbf{k}')i\sigma_2 \right ]^*_{s_1's_2'}+v_{t,\Gamma}\left[\hat{\mathbf{d}}_{\Gamma_j}(\mathbf{k})\cdot\boldsymbol{\sigma}i\sigma_2 \right ]_{s_1s_2}\left[\hat{\mathbf{d}}_{\Gamma_j}(\mathbf{k}')\cdot\boldsymbol{\sigma}i\sigma_2 \right ]^*_{s_1's_2'} \right ).
\label{eq:interaction_channels}
\end{equation}
\normalsize
Here, $j$ labels a basis function of an irreducible representation and $v_{s(t),\Gamma}$ are the interactions in the singlet (triplet) channels. Depending on the nature of the interactions, $v_{s(t),\Gamma}$ can be attractive (negative) or repulsive (positive).  
It is conceivable that a system close to a ferromagnetic instability would have the
spin-fluctuations opening an attractive triplet channel, with the electron-phonon interaction simultaneously generating an attractive singlet channel.
An explicit model of how different coexisting pairing mechanisms populate the pairing channels in Eq. \eqref{eq:interaction_channels} remains to be developed. 
The effects of attraction and repulsion on the singlet-triplet conversion process are studied in Refs. \cite{ PhysRevB.99.180505,ising_eilenberger_2020}. 
In noncentrossymetric materials, an additional singlet-triplet mixed channel is also allowed in Eq. \eqref{eq:interaction_channels}. Such a channel could be generated through Dzyaloshinskii-Moriya type interactions and would naturally lead to singlet-triplet mixing at zero field \cite{Frigeri2006}. Then why didn't we include it in Eq. \eqref{eq:interaction_channels}? We deliberately omit this term to show that singlet-triplet mixing can arise through the conversion mechanism in Eq. \eqref{eq:conversion} at finite fields.

\section{Locally non-centrosymmetric superconductors}

Materials that have an inversion center can exhibit properties of non-centrosymmetric materials. This can occur in layered crystals, where each layer alone lacks inversion, but when combined with the other layers, inversion symmetry is restored. 
Consider a system where there are two
inversion-related sublattices in the unit cell, which is the additional DOF for this material \cite{PRR2021aline,mockli2021}. 
The minimal Bogoliubov-deGennes (BdG) Hamiltonian is
\begin{equation}
\mathsf{H}_\mathsf{BdG} (\mathbf{k}) = \boldsymbol{\Psi}^\dag_\mathbf{k} 
\begin{bmatrix}
\hat{H}_0(\mathbf{k}) & \hat{\Delta}(\mathbf{k})\\ 
\hat{\Delta}^\dag(\mathbf{k}) & -\hat{H}_0^*(-\mathbf{k})
\end{bmatrix}
\boldsymbol{\Psi}_\mathbf{k},\quad \boldsymbol{\Psi}^\dag_\mathbf{k} = (\boldsymbol{\Phi}_\mathbf{k}^\dag,\boldsymbol{\Phi}^\mathsf{T}_{-\mathbf{k}}),\quad
\boldsymbol{\Phi}^\dag_\mathbf{k} = (c^\dag_{1\mathbf{k}\uparrow},c^\dag_{1\mathbf{k}\downarrow},c^\dag_{2\mathbf{k}\uparrow},c^\dag_{2\mathbf{k}\downarrow} ).
\label{eq:hbdg}
\end{equation}
Here, $c^\dag_{1\mathbf{k}\uparrow}$ is a creation operator of an electron in sublattice 1 with momentum $\mathbf{k}$ and spin $\uparrow$.
Using a basis in sublattice $\otimes$ spin space ($\tau_i\otimes\sigma_j$),
the normal and superconducting parts are described by
\begin{equation}
\hat{\mathsf{H}}_0(\mathbf{k})=\xi(\mathbf{k})\tau_0\otimes\sigma_0+\mathbf{t}(\mathbf{k})\cdot\boldsymbol{\tau}\otimes\sigma_0+\left[\tau_3\otimes\boldsymbol{\gamma}(\mathbf{k})-\tau_0\otimes\mathbf{B} \right ]\cdot\boldsymbol{\sigma},\quad
\hat{\Delta}(\mathbf{k})=\sum_{a,b=0}^3 \eta_{ab}\hat{d}_{ab}(\mathbf{k})\tau_a\otimes\sigma_b\,i\sigma_2. \label{eq:h0}
\end{equation}
Here, $\xi(\mathbf{k})$ comes from intra-sublattice hoppings, the vector $\mathbf{t}(\mathbf{k})$ describes contains the allowed inter-sublattice terms, $\boldsymbol{\gamma}(\mathbf{k})=(\gamma_x(\mathbf{k}),\gamma_y(\mathbf{k}),\gamma_z(\mathbf{k}))$ is a local antisymmetric spin-orbit coupling (SOC) that arises due to local inversion symmetry breaking, but global inversion exists due to $\tau_3$. 
The Zeeman magnetic field is $\mathbf{B}$. The sixteen possibilities of superconducting order parameters can be parametrized in the sublattice-spin space by $a,b$. In this basis, the $\eta_{30}$ order parameter corresponds to the state-vector with $c_7$ in Eq. \eqref{eq:odd}. 

\subsection{CeRh$_2$As$_2$}

We now apply the general ideas discussed above to the new two-phase superconductor CeRh$_2$As$_2$ \cite{khim2021fieldinduced}. 
The solid line in Fig. \ref{fig:1}(c) shows the normal to superconducting transition for perpendicular magnetic fields. 
From crystal structure, one expects that the in-plane Rashba SOC components $\{\gamma_x,\gamma_y\}$ are larger than the perpendicular Ising component $\gamma_z$ \cite{PRR2021aline}. Therefore, as a first approximation, we can consider the magnetic field $\mathbf{B}\perp\boldsymbol{\gamma}$.

A superconducting instability analysis shows that the state-vector $c_1$ in Eq. \eqref{eq:even} has the highest critical temperature at zero field \cite{PRR2021aline}. 
Since the crystal contains the inversion element, all odd states in Eq. \eqref{eq:odd} are prohibited to coexist with the even states in Eq. \eqref{eq:even}. 
However, an odd state-vector might switch with the initially dominant even state at high Zeeman fields.  
Interestingly, the odd-singlet $c_7$ state becomes energetically favorable at high magnetic fields. This happens because the $c_7$ states accommodates better to the joint presence of magnetic field, spin-orbit coupling and inter-sublattice hoppings. This is a universal property of locally non-centrosymmetric superconductors, and perhaps one can speculate more discoveries of dominant $c_7$ states in the near future. 

Because of the joint presence of spin-orbit coupling $\boldsymbol{\gamma}$ and magnetic field $\mathbf{B}\perp\boldsymbol{\gamma}$, if there is (even small) attraction in the triplet channels, the field converts $c_7$ singlets to equal spin $\{c_{11},c_{12}\}$ triplets. But this is not all. In the odd-phase, an additional conversion mechanism due to the joint presence of inter-sublattice hoppings $\mathbf{t}$ and magnetic field $\mathbf{B}$ also converts the $c_7$ singlets to inter-sublattice $\{c_{13},c_{14},c_{15},c_{16} \}$ triplets.
The zero-field $\{c_9,c_{10}\}$ triplets would require an intrinsic parity-mixed channel in the pairing interaction. 

\subsection{Bilayer-NbSe$_2$}

Transition metal dichalcogenide bilayers with $D_{3d}$ crystal symmetry are described by the same terms in Eq. \eqref{eq:h0}, but instead of Rashba SOC, one now has Ising SOC $\boldsymbol{\gamma}(\mathbf{k})=(0,0,\gamma_z(\mathbf{k}),0)$.
Since $\gamma_z$ locks the spins in the $z$-direction, these so called \textit{Ising superconductors} are robust against in-plane fields $\mathbf{B}\perp\boldsymbol{\gamma}$. 
The dominant superconducting component at zero field is very likely to be the $c_1$ state in Eq. \eqref{eq:even} \cite{kuzmanovic}. 
For in-plane fields, the same superconducting solutions found for CeRh$_2$As$_2$ for e perpendicular magnetic field apply to bilayer-TMDs. Also, because of the thin bilayer, an in-plane magnetic field couples dominantly to the spin degrees of freedom, such that orbital effects are expected to be very small. This makes bilayer-NbSe$_2$ a strong candidate for the $c_7$ odd-singlet state at high in-plane fields. 
It is possible that recent high field measurements already hint the $c_7$ state \cite{kuzmanovic}, but more high-field experimental data is needed. 

\section{Conclusion}

If additional DOFs lack, then singlet-triplet superconductivity can only occur via parity-mixing, which requires broken inversion symmetry. With additional DOFs, singlet-triplet pairing can occur within the same parity sector. The odd-parity sector contains only one singlet state. It is a universal property of locally non-centrosymmetric superconductors that the odd-parity singlet becomes favorable at high magnetic fields. If subleading pairing channels exist, the magnetic field converts the odd-singlets to both inter-sublattice triplets and intra-sublattice equal spin triplets.

\section*{References}
\bibliography{bib}

\end{document}